\documentclass[twocolumn,showpacs,preprintnumbers,amsmath,amssymb]{revtex4}

\usepackage{graphicx}
\usepackage{dcolumn}
\usepackage{bm}

\begin{document}

\preprint{APS/123-QED}

\title{Ab-initio calculation of the Gilbert damping parameter via linear
response formalism }

\author{H.~Ebert}
\author{S.~Mankovsky}
\author{D.~K\"odderitzsch}
\affiliation{
University of Munich,  
Department of Chemistry, 
Butenandtstrasse 5-13, D-81377 Munich, Germany
}%
\author{P.~J.~Kelly}
\affiliation{
Faculty of Science and Technology and MESA+ Institute for Nanotechnology,
University of Twente, P.O. Box 217, 7500 AE Enschede, The Netherlands
}%

\date{\today}

\begin{abstract}
A  Kubo-Greenwood-like  equation for the Gilbert damping
parameter $\alpha$  is presented that  is based  on the
linear response formalism. Its  implementation using
 the fully relativistic Korringa-Kohn-Rostoker
 (KKR) band structure method in combination with 
Coherent Potential Approximation (CPA) alloy theory 
allows it to be applied to a wide range of situations. This is
 demonstrated with results obtained for the bcc alloy system
Fe$_x$Co$_{1-x}$  as well as for a series of alloys  of permalloy  
with 5d transition metals.
 To account for the thermal displacements of atoms as a 
scattering mechanism, an alloy-analogy model is introduced.
The corresponding calculations for Ni correctly describe the rapid
change of $\alpha$ when small amounts of substitutional Cu are
introduced. 
\end{abstract}

\pacs{Valid PACS appear here}
\maketitle

\section{\label{sec:level1} Introduction }

 The magnetization dynamics that is relevant for 
the performance of any type
of magnetic device is in general governed by damping.
 In most cases the magnetization dynamics 
can be modeled successfully by means of the
 Landau-Lifshitz-Gilbert (LLG) equation \cite{Gil04}
that accounts for damping in a phenomenological way. 
The possibility to calculate the
corresponding damping parameter 
from first principles would open the perspective of optimizing materials
for devices and has therefore motivated extensive theoretical work in
the past. 
This led among others to Kambersky's breathing Fermi surface
 (BFS) \cite{Kam70} and torque-correlation model (TCM) \cite{Kam76}, 
that in principle provide a firm basis for numerical investigations
based on electronic structure calculations \cite{GIS07,FS06}.
The spin-orbit coupling that is seen as a key factor in transferring
energy from the magnetization to the electronic degrees of freedom is
explicitly included in these models. 
Most ab-initio results have been obtained for the BFS model though the
torque-correlation model makes fewer approximations \cite{GIS07,
  Kam07}.
In particular, it in principle describes the physical processes
responsible for Gilbert damping over a wide range of temperatures as
well as chemical (alloy) disorder. 
However, in practice, like many other models it depends on a relaxation
time parameter $\tau$ that 
describes the rate of transfer due to the various types of possible scattering 
mechanisms. This weak point could be removed  recently by
Brataas et al.\ \cite{BTB08}  who described the Gilbert
 damping by means of scattering theory. This development 
supplied the formal basis for the first parameter-free 
investigations on disordered alloys for which the dominant
 scattering mechanism is potential scattering caused
by chemical disorder \cite{SKB+10}. 

As pointed out by 
Brataas et al.\ \cite{BTB08}, their approach is completely 
equivalent to a formulation in terms of the linear response or Kubo
formalism. 
 The latter route is taken in this communication
 that presents a Kubo-Greenwood-like expression for the Gilbert damping
parameter. Application of the scheme to disordered 
alloys demonstrates that this approach is indeed fully equivalent
 to the scattering theory formulation of
Brataas et al.\ \cite{BTB08}. In addition a scheme is
 introduced to deal with the temperature dependence of 
the Gilbert damping parameter. 

\medskip

Following Brataas et al.\ \cite{BTB08}, 
the starting point of our scheme is the Landau-Lifshitz-Gilbert (LLG) 
equation  for the  time derivative  of the magnetization $\vec M$:
%
\begin{eqnarray}
\frac{1}{\gamma}\frac{d\vec{M}}{d\tau}  &= & 
-\vec{M}\times\vec{H}_{\rm eff}
+ \vec{M} \times  \left[ \frac{\tilde{G}(\vec{M})}{\gamma^2 M_s^2}
\frac{d\vec{M}}{d\tau} \right] \;,
\label{LLG}
\end{eqnarray}
%
 where  $ M_s$ is the saturation magnetization, $\gamma$  
the gyromagnetic ratio and $  \tilde G $ the Gilbert damping tensor.
 Accordingly, the time derivative of the magnetic energy is given by:
%
\begin{eqnarray}
\dot{E}_{\rm mag} = \vec{H}_{\rm eff}\cdot\frac{d\vec{M}}{d\tau} 
= \frac{1}{\gamma^2}\dot{\vec{m}}[\tilde{G}(\vec{m})\dot{\vec{m}}]
\label{MagnE}
\end{eqnarray}
%
in terms of the normalized magnetization $\vec{m} = \vec{M}/M_s$.
On the other hand the energy dissipation of the electronic system
$\dot{E}_{\rm dis} = \left\langle \frac{d\hat{H}}{d\tau}\right\rangle$ is
determined by the underlying  Hamiltonian $\hat{H}(\tau)$. Expanding the
normalized magnetization $\vec m(\tau)$, that determines the time dependence of  $\hat{H}(\tau)$ about its equilibrium value, 
$\vec{m}(\tau) = \vec{m}_0 + \vec{u}(\tau)$, one has:
%
\begin{eqnarray}
\hat{H} = \hat{H}_{0}(\vec{m}_0) + \sum_\mu \vec{u}_\mu\frac{\partial}{\partial \vec{u}_\mu} \hat{H}(\vec{m}_0) 
\; .
\label{E_expansion}
\end{eqnarray}
%
 Using the linear response formalism, $\dot{E}_{\rm dis}$ can be written as \cite{BTB08}:
%
\begin{eqnarray}
\dot{E}_{\rm dis} &=& -\pi\hbar \sum_{ii'}\sum_{\mu\nu}
\dot{u}_\mu \dot{u}_\nu
\left\langle  \psi_{i}|  \frac{\partial \hat{H}}{\partial u_{\mu}} | \psi_{i'}\right\rangle
\left\langle  \psi_{i'}| \frac{\partial \hat{H}}{\partial u_{\nu}}
  | \psi_i\right\rangle 
\nonumber \\
&&\;\;\;\;\;\;\;\;\;\;\;\;\;\;\;\;\;\;\times \delta(E_F - E_i)\delta(E_F - E_{i'})
\; ,
\label{E_dot}
\end{eqnarray}
%
where $ E _F$ is the Fermi energy and the sums run
over all eigenstates $ \alpha $ of the system.
 Identifying $\dot{E}_{\rm mag} = \dot{E}_{\rm dis}$, 
one gets an explicit expression for the 
Gilbert damping tensor $  \tilde G $ 
or equivalently for the  damping parameter 
 $\alpha = \tilde G /(\gamma M_s)$:
%
\begin{eqnarray}
\alpha_{\mu\nu} &=& -\frac{\pi\hbar \gamma}{M_s} \sum_{ii'}
\left\langle  \psi_{i}|  \frac{\partial \hat{H}}{\partial u_{\mu}} | \psi_{i'}\right\rangle
\left\langle  \psi_{i'}| \frac{\partial \hat{H}}{\partial u_{\nu}} | \psi_i\right\rangle
\nonumber  \\
&&\;\;\;\;\;\;\;\;\;\;\;\;\;\;\;\;\;\;\times \delta(E_F - E_i)\delta(E_F - E_{i'})
\; .
\label{alpha}
\end{eqnarray}
%
An efficient way to deal with Eq.\ (\ref{alpha}) is achieved by 
expressing the sum over the eigenstates by
means of the retarded single-particle Green's function 
 $\mbox{Im} G^{+}(E_F) = -\pi
\sum_{\alpha} |\psi_{\alpha}\rangle\langle\psi_{\alpha}|\delta(E_F - E_\alpha)$. 
This
leads for the parameter $\alpha $ to a Kubo-Greenwood-like equation:
%
\begin{eqnarray}
\alpha_{\mu\nu} &=& -\frac{\hbar \gamma}{\pi M_s} 
\mbox{Trace}
\left\langle  \frac{\partial \hat{H}}{\partial u_{\mu}} \mbox{Im}\; G^{+}(E_F)
\frac{\partial \hat{H}}{\partial u_{\nu}}  \mbox{Im}\; G^{+}(E_F)
\right\rangle_{c} 
\; 
\label{alpha2}
\end{eqnarray}
%
with $\langle ... \rangle_{c}$ indicating a configurational average in
case of a disordered system (see below).
Identifying $\partial \hat{H}/\partial u_\mu$ with the magnetic
torque $T_{\mu}$ this expression 
obviously gives the parameter $ \alpha $ in terms
 of a torque-torque correlation function. However, in contrast to the conventional 
TCM the electronic structure is not represented in terms of Bloch states 
but using the retarded electronic Green function giving the present
 approach much more flexibility. As it corresponds one-to-one to the 
standard Kubo-Greenwood equation for the electrical 
conductivity, the techniques developed to calculate conductivities can
be straightforwardly adopted to evaluate Eq.\ (\ref{alpha2}).

The most reliable way to account for spin-orbit coupling as the source
of Gilbert damping is to evaluate Eq.\ (\ref{alpha2}) using a fully
relativistic Hamiltonian within the framework of local spin density
formalism (LSDA) \cite{Ebe00}: 
%
\begin{eqnarray}
\hat{H} = c\vec{\alpha}\vec{p} +\beta  m c^2 +
 V(\vec r) + \beta \vec{\sigma}\vec{m}B(\vec r)
\; .
\label{Hamiltonian}           
\end{eqnarray}
%
Here $\alpha_i$ and $\beta$  are the standard Dirac matrices
 and $\vec p $
is the relativistic momentum operator \cite{Ros61}. 
The functions $V$ and
$B$  are the spin-averaged and spin-dependent  parts respectively of the
LSDA potential. 
 Eq.\ (\ref{Hamiltonian})  implies for the  magnetic
torque  $T_\mu$  occurring in
 Eq.\ (\ref{alpha2}) the expression: 
%
\begin{eqnarray}
T_\mu = \frac{\partial}{\partial u_\mu} \hat{H} = \beta B \sigma_\mu
\; .
\label{torque}
\end{eqnarray}
%
 The Green's function $G^{+} $ in Eq.\ (\ref{alpha})
 can be obtained in a very efficient way by using the spin-polarized  
 relativistic version of multiple
 scattering theory \cite{Ebe00} that allows us to treat magnetic solids: 
%
\begin{eqnarray}
 G^{+}(\vec{r}_n,\vec r_m\,',E)& = &
 \sum_{\Lambda \Lambda'}
Z^{n}_{\Lambda}(\vec r_n,E)
\,
  {\tau}_{ \Lambda  \Lambda'}^{nm}(E)
\,
Z^{m\times}_{\Lambda'}(\vec r_m\,',E)
\nonumber \\
&&
- \sum_{\Lambda }
Z^{n}_{\Lambda}(\vec r_<,E)
\,
J^{n\times}_{\Lambda'}(\vec r_>,E)
\,\delta_{nm}
\; .
\label{GreensFunction}
\end{eqnarray}
%
Here coordinates $\vec{r}_n$ referring to the center of cell $n$ have
been used with  $|\vec{r}_<| = min(|\vec{r}_n|,|\vec{r}_n\,'|)$ and
$|\vec{r}_>| = max(|\vec{r}_n|,|\vec{r}_n\,'|)$.
The four component wave functions $Z^{n}_{\Lambda}(\vec r,E)$ 
($J^{n}_{\Lambda}(\vec r,E)$) are
regular (irregular)
 solutions to the single-site Dirac equation  for site $n$ and 
${\tau}^{nm}_{ \Lambda  \Lambda'} (E)$ is
 the so-called scattering path operator
 that transfers an electronic wave coming in at 
site $ m $ into a wave going out from site $ n $ with
all possible intermediate scattering events accounted for coherently. 

 Using matrix notation, this
leads to the following expression for the damping parameter:
\begin{eqnarray}
\alpha_{\mu\mu} =   \frac{g}{\pi\mu_{tot}} \sum_{n } \mbox{ Trace}
\left\langle \underline{T}^{0\mu} \,
 \tilde{\underline{\tau}}^{0n}\,
\underline{T}^{n\mu} \,
 \tilde{\underline{\tau}}^{n0} \right\rangle_{c}
\label{alpha_MST}
\end{eqnarray}
%
with the g-factor $2(1 + {\mu_{orb}}/{\mu_{spin}})$ in terms of
the spin and orbital moments, $\mu_{spin}$ and $\mu_{orb}$,
respectively, the total magnetic moment $\mu_{tot} =
\mu_{spin}+\mu_{orb}$, and
$\tilde{\tau}_{\Lambda\Lambda'}^{0n} =
\frac{1}{2i}(\tau_{\Lambda\Lambda'}^{0n} -
  \tau_{\Lambda'\Lambda}^{0n})$ and the energy argument $E_F$
omitted.
 The matrix elements of the torque operator $T^{n\mu}$  are identical to
those occurring in the context of exchange coupling  \cite{EM09a}  and can be expressed in
 terms of the spin-dependent part $B$ of the electronic potential with matrix elements:
%

\begin{eqnarray}
T_{\Lambda'\Lambda}^{n\mu} & = & \int d^3r\; Z^{n\times}_{\Lambda'}(\vec{r})\;\left[\beta
\sigma_{\mu}B_{xc}(\vec{r})\right] Z^{n}_{\Lambda}(\vec{r}) 
\label{matrix-element}
\; .
\end{eqnarray}
%

 As indicated above, the expressions in 
Eqs.\  (\ref{alpha2}) -- (\ref{matrix-element})
can be applied straightforwardly to disordered alloys. 
In this case the brackets
 $\left\langle  ...\right\rangle_{c}  $
 indicate the necessary configurational average. 
This can be done by describing in a first step 
the underlying electronic structure (for $T=0 $~K)
on the basis of the Coherent Potential Approximation (CPA) alloy theory.
 In the next step the configurational average in Eq. (\ref{alpha2})
 is taken following the 
scheme worked out by Butler \cite{But85}
 when dealing with the electrical conducting at $T=0 $~K
 or residual resistivity respectively, of disordered alloys.
This implies in particular that so-called vertex corrections of the type
 $\left\langle  T_{\mu} \mbox{Im} G^{+} T_{\nu} \mbox{Im} G^{+}
\right\rangle_{c} -
\left\langle  T_{\mu} \mbox{Im} G^{+}\right\rangle_{c}
\left\langle  T_{\nu} \mbox{Im} G^{+}\right\rangle_{c}
$
 that account for scattering-in processes in the language of the 
Boltzmann transport formalism are properly accounted for.

Thermal vibrations as a source of electron scattering can in principle
be accounted for by a generalization of Eqs.\  (\ref{alpha2}) --
(\ref{matrix-element}) to finite temperatures and by including the
electron-phonon self-energy $\Sigma_{el-ph}$ when calculating the Greens
function $G^+$. Here we restrict ourselves to elastic scattering
processes by using a quasi-static representation of the thermal
displacements of the atoms 
from their equilibrium positions. 
 We introduce an alloy-analogy model
 to average over a discrete set of displacements that is 
chosen to reproduce the thermal root mean square average displacement
$\sqrt{\langle u^2\rangle_T}$  for a given temperature $T $. 
This was chosen according to ${\langle u^2\rangle_T} =
\frac{1}{4}\frac{3h^2}{\pi^2mk\Theta_D}[\frac{\Phi(\Theta_D/T)}{\Theta_D/T}
+\frac{1}{4}]$ with 
$\Phi(\Theta_D/T)$ the Debye function,
$h$ the Planck constant, $k$ the Boltzmann constant and
$\Theta_D$ the Debye temperature \cite{GMMP83}.
Ignoring the zero temperature term $1/4$ and
assuming a frozen potential for
the atoms, the situation can be dealt with in full
analogy to the treatment of disordered alloys described above.

\begin{figure}
  \begin{center}
 \includegraphics[scale=0.4]{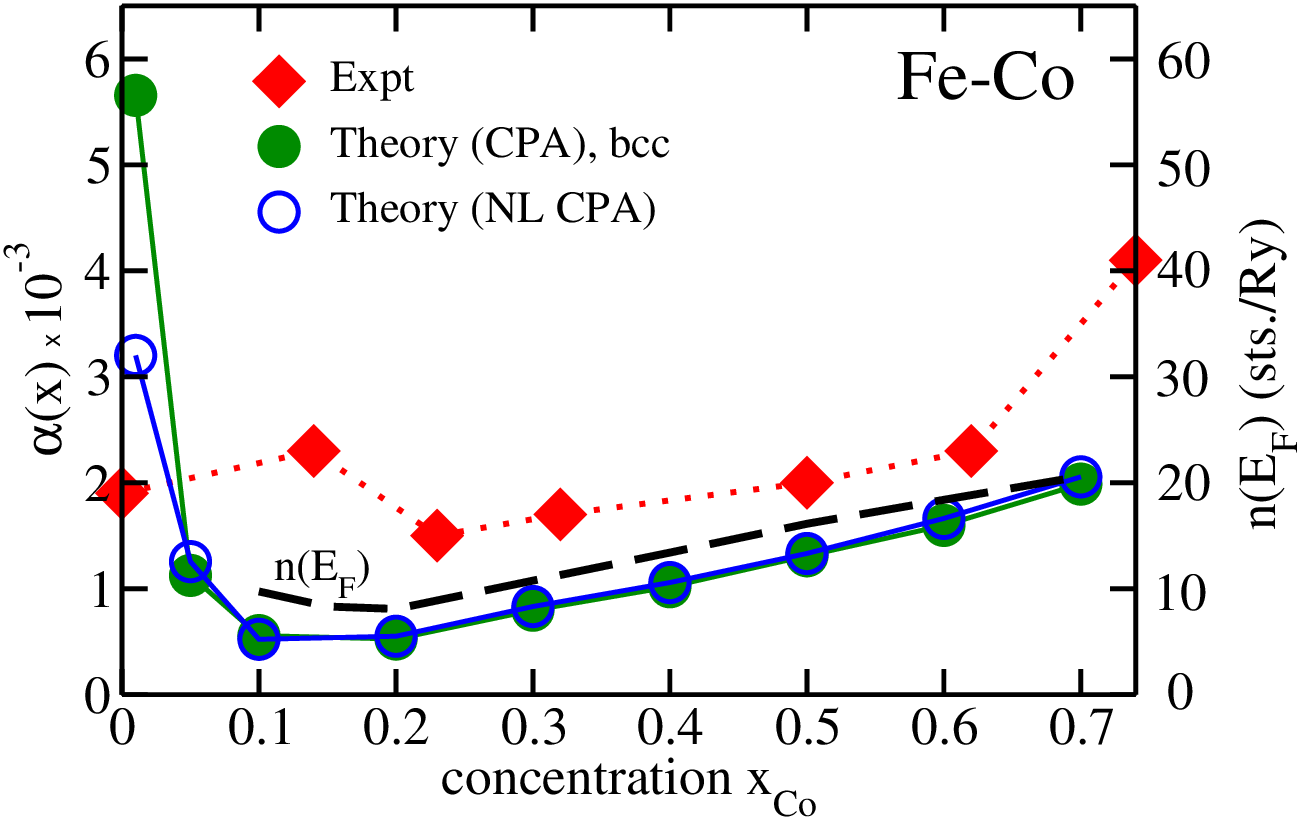}
  \end{center}
  \vskip-5mm
  \caption{\label{Fig-FeNi}
Gilbert damping parameter for bcc Fe$_x$Co$_{1-x}$ as a function
    of Co concentration: full circles - the present results within CPA, empty
    circles - within non-local CPA (NL CPA), and full diamonds -
    experimental data by Oogane \cite{OWY+06}.}
  \vskip-6mm  
\end{figure}

 The approach described above has been applied to the 
ferromagnetic 3d-transition metal alloy systems
 bcc Fe$_x$Co$_{1-x}$,  fcc Fe$_x$Ni$_{1-x}$  and
 fcc Co$_x$Ni$_{1-x}$. 
Fig.\ \ref{Fig-FeNi}  shows as an example results for
  bcc Fe$_x$Co$_{1-x}$   for $x \le 0.7$.
 The calculated damping parameter $ \alpha (x) $ for  $T=0 $~K
 is found in very good agreement with the results 
based on the scattering theory approach \cite{SKB+10}
demonstrating  numerically the equivalence of the two approaches. 
An indispensable requirement to achieve this agreement is to
 include the vertex corrections mentioned above. In fact, ignoring them 
leads in some cases to completely unphysical results. 
To check the reliability of the standard CPA, that implies a single-site
approximation when performing the configurational average, we performed
calculations on the basis of the non-local CPA \cite{KERE07}. In this
case four atom cluster have been used leading - apart from the very
dilute case - practically to the same results as the CPA. 
As found 
before  for fcc Fe$_x$Ni$_{1-x}$ \cite{SKB+10} the theoretical results
for $\alpha$ reproduce the
concentration dependence of the experimental data quite well but are
found too low (see below).
As suggested by Eq. (\ref{alpha_MST}) the variation of $\alpha(x)$ with
concentration $x$ may reflect to some extent the variation of the
average magnetic moment $\mu_{tot}$ of alloy. As the moments as well as
the spin-orbit coupling strength of Fe and Co don't differ too much, the
variation of $\alpha(x)$ should be determined in the concentrated regime
primarily by the electronic structure at the Fermi energy $E_F$. As
Fig. \ref{Fig-FeNi} shows, there is indeed a close correlation of the
density of states $n(E_F)$ that may be seen as a measure for the
available relaxation channels.

\begin{figure}[hbt]
  \begin{center}
 \includegraphics[scale=0.58]{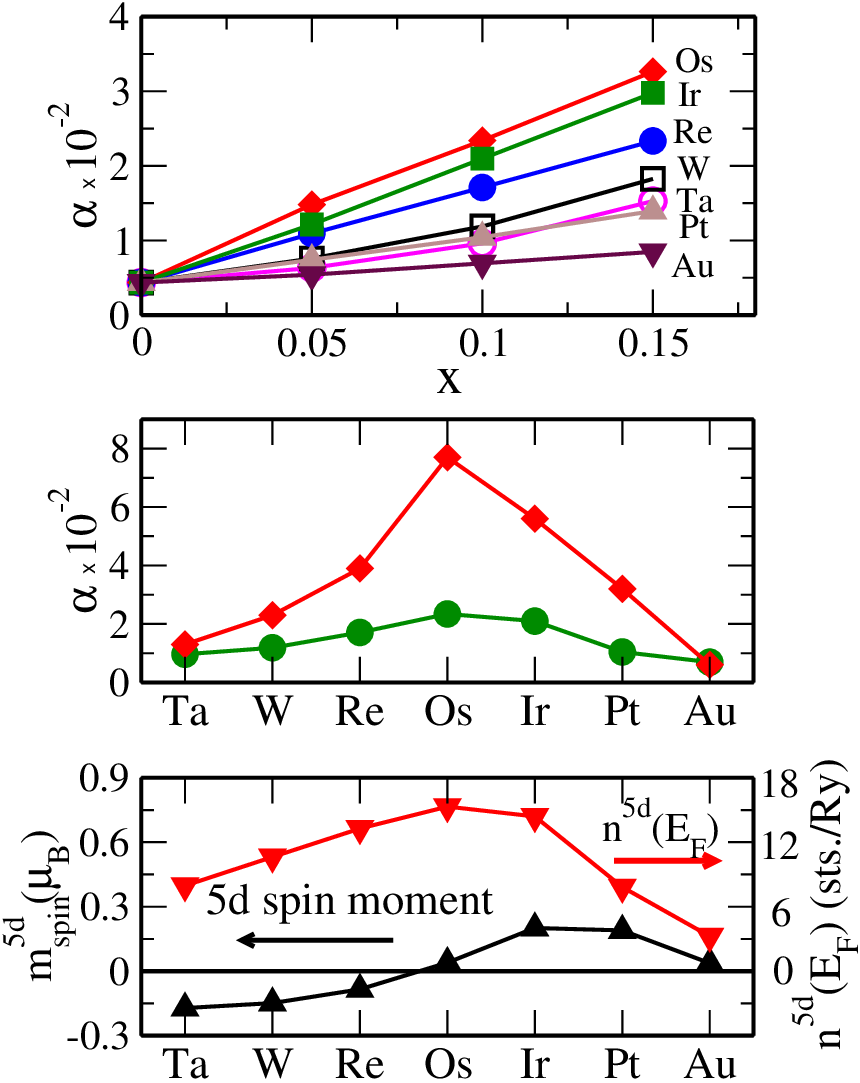}
   \end{center}
  \vskip-5mm
  \caption{\label{Fig-Py5d}
Top:  Change of the Gilbert damping parameter  $\Delta \alpha$  w.r.t.\ 
permalloy (Py) for various Py/5d TM systems as a function of 5d  TM
concentration; 
Middle: Gilbert damping parameter $\alpha$ for Py/5d TM systems 
   with 10 \% 5d TM  content in comparison
with  experiment  \cite{RMC+07}; 
Bottom: spin magnetic moment $m_{spin}^{5d}$ and density of states $n(E_F)$
at the Fermi energy of the $5d$ component in Py/5d TM systems 
   with 10 \% 5d TM content.  }
  \end{figure}

 While the scattering and linear response approach are completely
 equivalent when dealing with bulk alloys the latter allows us to
 perform the necessary configuration averaging in a much more
 efficient way. This allows us to study with moderate effort the
 influence of varying the alloy composition on the damping
 parameter $ \alpha $. Corresponding work has been done in
 particular using permalloy as a starting material and adding 
transition metals (TM) \cite{RMC+07}  or rare earth metals \cite{WKM+09}.
Fig.\ \ref{Fig-Py5d}  (top) shows results obtained by substituting 
Fe and Ni atoms in permalloy by 5d TMs. 
As found  by experiment \cite{RMC+07} $\alpha $
 increases in all cases nearly linearly
 with  the 5d TM content. The total damping
 for 10 \% 5d TM  content  shown in the middle panel of
Fig.\ \ref{Fig-Py5d}  varies roughly parabolically
 over the 5d TM series. 
In contrast to the Fe$_x$Co$_{1-x}$ alloys considered above, 
there is now an S-like variation of the moments $\mu_{spin}^{5d}$ over the
    series (Fig.  \ref{Fig-Py5d}, bottom), characteristic of 5d
    impurities in the pure hosts Fe and Ni \cite{DSB+89,SOZD87}.
In spite of this behaviour of $\mu_{spin}^{5d}$
the variation of $\alpha(x)$ seems again to be correlated
with the density of states $n^{5d}(E_F)$ (Fig. \ref{Fig-Py5d} bottom).
Again the trend of the experimental
data is well reproduced by the theoretical ones
 that are however 
somewhat too low.

\begin{figure}
  \begin{center}
 \includegraphics[scale=0.6]{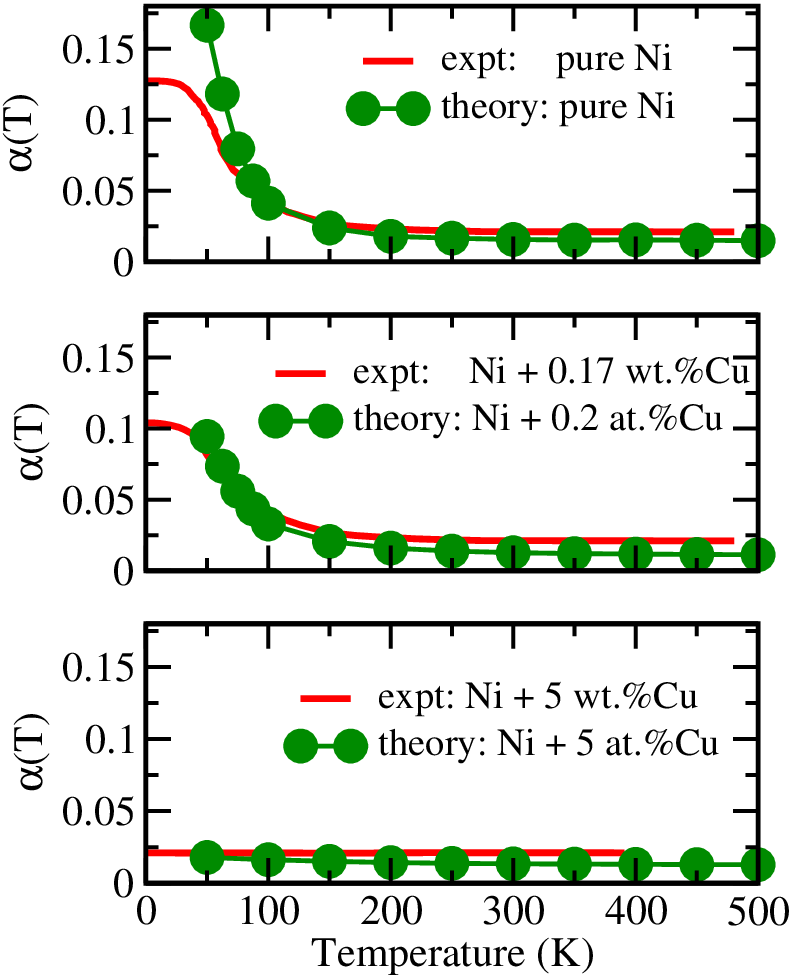}
  \end{center}
  \vskip-5mm
  \caption{ \label{Fig-Ni}
Temperature variation of Gilbert damping of pure Ni and Ni
    with Cu impurities: present theoretical results vs experiment \cite{BL74}}
  \vskip-6mm
 \end{figure}

 One of the possible reasons for the discrepancy of
 the theoretical and experimental results shown in
 Figs.\  \ref{Fig-FeNi}  and \ref{Fig-Py5d}  might be
 the neglect of the influence of finite temperatures.
 This can be incorporated as indicated above by
 accounting for the thermal displacement of the atoms 
in a quasi-static way and performing  a configurational 
average over the displacements using the CPA.
 This leads even for pure systems to a scattering mechanism 
and this way to a finite value for $ \alpha $.
 Corresponding results for pure Ni are given in
Fig.\  \ref{Fig-Ni}  that  show in full accordance 
with experiment a rapid decrease 
of $\alpha $ with increasing temperature until 
a regime with a weak variation of $\alpha$ with $T$
is reached. 
 This behavior is 
commonly interpreted as a transition 
from conductivity-like to resistivity-like behaviour
reflecting the dominance of intra- and inter-band
transition, respectively \cite{GIS07},
that is related to the increase of the broadening of electron energy
bands caused by the increase of scattering events with temperature.
 Adding only less than 1 at.\ \% Cu to Ni, the 
conductivity-like behavior at low temperatures
 is strongly reduced while the high temperature
behavior is hardly changed.  A further increase of
 the Cu content
leads to the impurity-scattering processes responsible for the band
broadening dominating $\alpha$. This effect
 completely suppresses the 
conductivity-like behavior in the low-temperature
 regime because of the increase of scattering
 events due to chemical disorder.
 Again this is fully in line with  the experimental  data,
providing a straightforward explanation for  their peculiar  variation 
with temperature and composition.

From the results obtained for Ni one may conclude that thermal lattice
displacements are only partly responsible for the finding 
 that the damping parameters obtained for Py doped with the $5d$ TM
    series, and Fe$_x$Co$_{1-x}$ are somewhat low compared with experiment. 
This indicates
that additional relaxation mechanisms like magnon scattering
contribute. Again, these can be included at least in a quasi-static way
by adopting the point of view of a disordered local moment picture. This
implies scattering due to random temperature-dependent fluctuations of
the spin moments that can also be dealt with using the CPA.

\medskip
 In summary, a  formulation for the Gilbert damping parameter
 $ \alpha $ in terms of a torque-torque-correlation function
 was derived that  led to a  Kubo-Greenwood-like equation. 
  The scheme was implemented using the fully relativistic 
KKR band structure method in combination with the CPA 
alloy theory. 
This allows us to account for various types of scattering mechanisms in
a parameter-free way. 
 Corresponding applications to disordered transition metal
 alloys led to very good agreement with results based on 
the scattering theory approach of Brataas et al.\ 
demonstrating the equivalence of both approaches.
 The flexibility and numerical efficiency of the present
 scheme was demonstrated by a study
 on a series of permalloy-5d TM systems. 
To investigate the influence of finite temperatures on
$\alpha $, a so-called 
alloy-analogy model was introduced that deals with the
 thermal displacement of atoms in a quasi-static manner.
 Applications to pure Ni gave results in very good agreement
with experiment and in particular reproduced the dramatic
change of $\alpha $ when Cu is added to Ni.

\begin{acknowledgments}

The authors would like to thank the DFG for 
financial support within
the SFB 689 ``Spinph\"anomene in reduzierten Dimensionen''  and within
project Eb154/23 for financial support.
PJK acknowledges support by EU FP7 ICT Grant No. 251759 MACALO.

\end{acknowledgments}




\begin{thebibliography}{20}
\expandafter\ifx\csname natexlab\endcsname\relax\def\natexlab#1{#1}\fi
\expandafter\ifx\csname bibnamefont\endcsname\relax
  \def\bibnamefont#1{#1}\fi
\expandafter\ifx\csname bibfnamefont\endcsname\relax
  \def\bibfnamefont#1{#1}\fi
\expandafter\ifx\csname citenamefont\endcsname\relax
  \def\citenamefont#1{#1}\fi
\expandafter\ifx\csname url\endcsname\relax
  \def\url#1{\texttt{#1}}\fi
\expandafter\ifx\csname urlprefix\endcsname\relax\def\urlprefix{URL }\fi
\providecommand{\bibinfo}[2]{#2}
\providecommand{\eprint}[2][]{\url{#2}}

\bibitem[{\citenamefont{Gilbert}(2004)}]{Gil04}
\bibinfo{author}{\bibfnamefont{T.~L.} \bibnamefont{Gilbert}},
  \bibinfo{journal}{IEEE Transactions on Magnetics}
  \textbf{\bibinfo{volume}{40}}, \bibinfo{pages}{3443} (\bibinfo{year}{2004}).

\bibitem[{\citenamefont{Kambersky}(1970)}]{Kam70}
\bibinfo{author}{\bibfnamefont{V.}~\bibnamefont{Kambersky}},
  \bibinfo{journal}{Can. J. Phys.} \textbf{\bibinfo{volume}{48}},
  \bibinfo{pages}{2906} (\bibinfo{year}{1970}).

\bibitem[{\citenamefont{Kambersky}(1976)}]{Kam76}
\bibinfo{author}{\bibfnamefont{V.}~\bibnamefont{Kambersky}},
  \bibinfo{journal}{Czech. J. Phys.} \textbf{\bibinfo{volume}{26}},
  \bibinfo{pages}{1366} (\bibinfo{year}{1976}),
  \urlprefix\url{http://dx.doi.org/10.1007/BF01587621}.

\bibitem[{\citenamefont{Gilmore et~al.}(2007)\citenamefont{Gilmore, Idzerda,
  and Stiles}}]{GIS07}
\bibinfo{author}{\bibfnamefont{K.}~\bibnamefont{Gilmore}},
  \bibinfo{author}{\bibfnamefont{Y.~U.} \bibnamefont{Idzerda}},
  \bibnamefont{and} \bibinfo{author}{\bibfnamefont{M.~D.}
  \bibnamefont{Stiles}}, \bibinfo{journal}{Phys. Rev. Lett.}
  \textbf{\bibinfo{volume}{99}}, \bibinfo{pages}{027204}
  (\bibinfo{year}{2007}),
  \urlprefix\url{http://link.aps.org/doi/10.1103/PhysRevLett.99.027204}.

\bibitem[{\citenamefont{F\"ahnle and Steiauf}(2006)}]{FS06}
\bibinfo{author}{\bibfnamefont{M.}~\bibnamefont{F\"ahnle}} \bibnamefont{and}
  \bibinfo{author}{\bibfnamefont{D.}~\bibnamefont{Steiauf}},
  \bibinfo{journal}{Phys. Rev. B} \textbf{\bibinfo{volume}{73}},
  \bibinfo{pages}{184427} (\bibinfo{year}{2006}).

\bibitem[{\citenamefont{Kambersky}(2007)}]{Kam07}
\bibinfo{author}{\bibfnamefont{V.}~\bibnamefont{Kambersky}},
  \bibinfo{journal}{Phys. Rev. B} \textbf{\bibinfo{volume}{76}},
  \bibinfo{pages}{134416} (\bibinfo{year}{2007}).

\bibitem[{\citenamefont{Brataas et~al.}(2008)\citenamefont{Brataas,
  Tserkovnyak, and Bauer}}]{BTB08}
\bibinfo{author}{\bibfnamefont{A.}~\bibnamefont{Brataas}},
  \bibinfo{author}{\bibfnamefont{Y.}~\bibnamefont{Tserkovnyak}},
  \bibnamefont{and} \bibinfo{author}{\bibfnamefont{G.~E.~W.}
  \bibnamefont{Bauer}}, \bibinfo{journal}{Phys. Rev. Lett.}
  \textbf{\bibinfo{volume}{101}}, \bibinfo{pages}{037207}
  (\bibinfo{year}{2008}),
  \urlprefix\url{http://link.aps.org/doi/10.1103/PhysRevLett.101.037207}.

\bibitem[{\citenamefont{Starikov et~al.}(2010)\citenamefont{Starikov, Kelly,
  Brataas, Tserkovnyak, and Bauer}}]{SKB+10}
\bibinfo{author}{\bibfnamefont{A.~A.} \bibnamefont{Starikov}},
  \bibinfo{author}{\bibfnamefont{P.~J.} \bibnamefont{Kelly}},
  \bibinfo{author}{\bibfnamefont{A.}~\bibnamefont{Brataas}},
  \bibinfo{author}{\bibfnamefont{Y.}~\bibnamefont{Tserkovnyak}},
  \bibnamefont{and} \bibinfo{author}{\bibfnamefont{G.~E.~W.}
  \bibnamefont{Bauer}}, \bibinfo{journal}{Phys. Rev. Lett.}
  \textbf{\bibinfo{volume}{105}}, \bibinfo{pages}{236601}
  (\bibinfo{year}{2010}),
  \urlprefix\url{http://link.aps.org/doi/10.1103/PhysRevLett.105.236601}.

\bibitem[{\citenamefont{Ebert}(2000)}]{Ebe00}
\bibinfo{author}{\bibfnamefont{H.}~\bibnamefont{Ebert}}, in
  \emph{\bibinfo{booktitle}{Electronic Structure and Physical Properties of
  Solids}}, edited by
  \bibinfo{editor}{\bibfnamefont{H.}~\bibnamefont{Dreyss\'{e}}}
  (\bibinfo{publisher}{Springer}, \bibinfo{address}{Berlin},
  \bibinfo{year}{2000}), vol. \bibinfo{volume}{535} of
  \emph{\bibinfo{series}{Lecture Notes in Physics}}, p. \bibinfo{pages}{191}.

\bibitem[{\citenamefont{Rose}(1961)}]{Ros61}
\bibinfo{author}{\bibfnamefont{M.~E.} \bibnamefont{Rose}},
  \emph{\bibinfo{title}{Relativistic Electron Theory}}
  (\bibinfo{publisher}{Wiley}, \bibinfo{address}{New York},
  \bibinfo{year}{1961}).

\bibitem[{\citenamefont{Ebert and Mankovsky}(2009)}]{EM09a}
\bibinfo{author}{\bibfnamefont{H.}~\bibnamefont{Ebert}} \bibnamefont{and}
  \bibinfo{author}{\bibfnamefont{S.}~\bibnamefont{Mankovsky}},
  \bibinfo{journal}{Phys. Rev. B} \textbf{\bibinfo{volume}{79}},
  \bibinfo{pages}{045209} (\bibinfo{year}{2009}),
  \urlprefix\url{http://link.aps.org/doi/10.1103/PhysRevB.79.045209}.

\bibitem[{\citenamefont{Butler}(1985)}]{But85}
\bibinfo{author}{\bibfnamefont{W.~H.} \bibnamefont{Butler}},
  \bibinfo{journal}{Phys. Rev. B} \textbf{\bibinfo{volume}{31}},
  \bibinfo{pages}{3260} (\bibinfo{year}{1985}),
  \urlprefix\url{http://link.aps.org/doi/10.1103/PhysRevB.31.3260}.

\bibitem[{\citenamefont{Gololobov et~al.}(1983)\citenamefont{Gololobov, Mager,
  Mezhevich, and Pan}}]{GMMP83}
\bibinfo{author}{\bibfnamefont{E.~M.} \bibnamefont{Gololobov}},
  \bibinfo{author}{\bibfnamefont{E.~L.} \bibnamefont{Mager}},
  \bibinfo{author}{\bibfnamefont{Z.~V.} \bibnamefont{Mezhevich}},
  \bibnamefont{and} \bibinfo{author}{\bibfnamefont{L.~K.} \bibnamefont{Pan}},
  \bibinfo{journal}{phys. stat. sol. (b)} \textbf{\bibinfo{volume}{119}},
  \bibinfo{pages}{K139} (\bibinfo{year}{1983}).

\bibitem[{\citenamefont{Oogane et~al.}(2006)\citenamefont{Oogane, Wakitani,
  Yakata, Yilgin, Ando, Sakuma, and Miyazaki}}]{OWY+06}
\bibinfo{author}{\bibfnamefont{M.}~\bibnamefont{Oogane}},
  \bibinfo{author}{\bibfnamefont{T.}~\bibnamefont{Wakitani}},
  \bibinfo{author}{\bibfnamefont{S.}~\bibnamefont{Yakata}},
  \bibinfo{author}{\bibfnamefont{R.}~\bibnamefont{Yilgin}},
  \bibinfo{author}{\bibfnamefont{Y.}~\bibnamefont{Ando}},
  \bibinfo{author}{\bibfnamefont{A.}~\bibnamefont{Sakuma}}, \bibnamefont{and}
  \bibinfo{author}{\bibfnamefont{T.}~\bibnamefont{Miyazaki}},
  \bibinfo{journal}{Jap. J. Appl. Phys.} \textbf{\bibinfo{volume}{45}},
  \bibinfo{pages}{3889} (\bibinfo{year}{2006}).

\bibitem[{\citenamefont{K\"odderitzsch
  et~al.}(2007)\citenamefont{K\"odderitzsch, Ebert, Rowlands, and
  Ernst}}]{KERE07}
\bibinfo{author}{\bibfnamefont{D.}~\bibnamefont{K\"odderitzsch}},
  \bibinfo{author}{\bibfnamefont{H.}~\bibnamefont{Ebert}},
  \bibinfo{author}{\bibfnamefont{D.~A.} \bibnamefont{Rowlands}},
  \bibnamefont{and} \bibinfo{author}{\bibfnamefont{A.}~\bibnamefont{Ernst}},
  \bibinfo{journal}{New Journal of Physics} \textbf{\bibinfo{volume}{9}},
  \bibinfo{pages}{81} (\bibinfo{year}{2007}),
  \urlprefix\url{http://dx.doi.org/10.1088/1367-2630/9/4/081}.

\bibitem[{\citenamefont{Rantschler et~al.}(2007)\citenamefont{Rantschler,
  McMichael, Castillo, Shapiro, Egelhoff, Maranville, Pulugurtha, Chen, and
  Connors}}]{RMC+07}
\bibinfo{author}{\bibfnamefont{J.~O.} \bibnamefont{Rantschler}},
  \bibinfo{author}{\bibfnamefont{R.~D.} \bibnamefont{McMichael}},
  \bibinfo{author}{\bibfnamefont{A.}~\bibnamefont{Castillo}},
  \bibinfo{author}{\bibfnamefont{A.~J.} \bibnamefont{Shapiro}},
  \bibinfo{author}{\bibfnamefont{W.~F.} \bibnamefont{Egelhoff}},
  \bibinfo{author}{\bibfnamefont{B.~B.} \bibnamefont{Maranville}},
  \bibinfo{author}{\bibfnamefont{D.}~\bibnamefont{Pulugurtha}},
  \bibinfo{author}{\bibfnamefont{A.~P.} \bibnamefont{Chen}}, \bibnamefont{and}
  \bibinfo{author}{\bibfnamefont{L.~M.} \bibnamefont{Connors}},
  \bibinfo{journal}{J. Appl. Phys.} \textbf{\bibinfo{volume}{101}},
  \bibinfo{pages}{033911} (\bibinfo{year}{2007}).

\bibitem[{\citenamefont{Woltersdorf et~al.}(2009)\citenamefont{Woltersdorf,
  Kiessling, Meyer, Thiele, and Back}}]{WKM+09}
\bibinfo{author}{\bibfnamefont{G.}~\bibnamefont{Woltersdorf}},
  \bibinfo{author}{\bibfnamefont{M.}~\bibnamefont{Kiessling}},
  \bibinfo{author}{\bibfnamefont{G.}~\bibnamefont{Meyer}},
  \bibinfo{author}{\bibfnamefont{J.-U.} \bibnamefont{Thiele}},
  \bibnamefont{and} \bibinfo{author}{\bibfnamefont{C.~H.} \bibnamefont{Back}},
  \bibinfo{journal}{Phys. Rev. Lett.} \textbf{\bibinfo{volume}{102}},
  \bibinfo{pages}{257602} (\bibinfo{year}{2009}),
  \urlprefix\url{http://link.aps.org/doi/10.1103/PhysRevLett.102.257602}.

\bibitem[{\citenamefont{Drittler et~al.}(1989)\citenamefont{Drittler, Stefanou,
  Bl\"ugel, Zeller, and Dederichs}}]{DSB+89}
\bibinfo{author}{\bibfnamefont{B.}~\bibnamefont{Drittler}},
  \bibinfo{author}{\bibfnamefont{N.}~\bibnamefont{Stefanou}},
  \bibinfo{author}{\bibfnamefont{S.}~\bibnamefont{Bl\"ugel}},
  \bibinfo{author}{\bibfnamefont{R.}~\bibnamefont{Zeller}}, \bibnamefont{and}
  \bibinfo{author}{\bibfnamefont{P.~H.} \bibnamefont{Dederichs}},
  \bibinfo{journal}{Phys. Rev. B} \textbf{\bibinfo{volume}{40}},
  \bibinfo{pages}{8203} (\bibinfo{year}{1989}),
  \urlprefix\url{http://link.aps.org/doi/10.1103/PhysRevB.40.8203}.

\bibitem[{\citenamefont{Stefanou et~al.}(1987)\citenamefont{Stefanou, Oswald,
  Zeller, and Dederichs}}]{SOZD87}
\bibinfo{author}{\bibfnamefont{N.}~\bibnamefont{Stefanou}},
  \bibinfo{author}{\bibfnamefont{A.}~\bibnamefont{Oswald}},
  \bibinfo{author}{\bibfnamefont{R.}~\bibnamefont{Zeller}}, \bibnamefont{and}
  \bibinfo{author}{\bibfnamefont{P.~H.} \bibnamefont{Dederichs}},
  \bibinfo{journal}{Phys. Rev. B} \textbf{\bibinfo{volume}{35}},
  \bibinfo{pages}{6911} (\bibinfo{year}{1987}),
  \urlprefix\url{http://link.aps.org/doi/10.1103/PhysRevB.35.6911}.

\bibitem[{\citenamefont{Bhagat and Lubitz}(1974)}]{BL74}
\bibinfo{author}{\bibfnamefont{S.~M.} \bibnamefont{Bhagat}} \bibnamefont{and}
  \bibinfo{author}{\bibfnamefont{P.}~\bibnamefont{Lubitz}},
  \bibinfo{journal}{Phys. Rev. B} \textbf{\bibinfo{volume}{10}},
  \bibinfo{pages}{179} (\bibinfo{year}{1974}).

\end{thebibliography}

\end{document}